\documentclass[journal=apchd5,manuscript=letter,amsmath,amssymb,]{achemso}

\usepackage{chemformula} 
\usepackage[T1]{fontenc} 

\newcommand{\beginsupplement}{%
        \setcounter{table}{0}
        \renewcommand{\thetable}{SI\arabic{table}}%
        \setcounter{figure}{0}
        \renewcommand{\thefigure}{S\arabic{figure}}%
        \setcounter{equation}{0}
        \renewcommand{\theequation}{S\arabic{equation}}%
        \setcounter{section}{0}
        \renewcommand{\thesection}{S\arabic{section}}%
        \setcounter{page}{1}
        \renewcommand{\thepage}{S\arabic{page}}%
     }

\usepackage{xcolor}
\usepackage[final]{changes}
\definechangesauthor[color=red, name={Paulo Santos}]{PVS}
\definechangesauthor[color=blue]{MY}
\newcommand{\rPVS}[2]{\replaced{#1}{#2}}
\newcommand{\aPVS}[1]{\added{#1}}
\newcommand{\dPVS}[1]{\deleted{#1}}
\newcommand{\rMY}[2]{\replaced{#1}{#2}}
\newcommand{\aMY}[1]{\added{#1}}
\newcommand{\dMY}[1]{\deleted{#1}}
\newcommand{\lSAW}{{\lambda_{\mathrm{SAW}}}}		
\newcommand{\vSAW}{v_\mathrm{SAW}}			
\newcommand{\fSAW}{f_\mathrm{SAW}}			
\newcommand{\TSAW}{T_\mathrm{SAW}}			
\newcommand{\Prf}{P_\mathrm{rf}} 			
\newcommand{\Pl}{P_\ell}				
\newcommand{\DB}{\mathrm{D_B}}

\author{Mingyun Yuan}
\email{yuan@pdi-berlin.de}
\author{Klaus Biermann}
\affiliation{Paul-Drude-Institut f{\"u}r Festk{\"o}rperelektronik, 
Leibniz-Institut im Forschungsverbund Berlin e.V., 
Hausvogteiplatz 5-7, 10117 Berlin, Germany}
\author{Shintaro Takada}
\affiliation{National Institute of Advanced Industrial Science and Technology (AIST), National Metrology Institute of Japan (NMIJ), 1-1-1 Umezono, Tsukuba, Ibaraki 305-8563, Japan}
\author{Christopher B\"auerle}
\affiliation{Univ. Grenoble Alpes, CNRS, Grenoble INP, Institut N{\'e}el, 38000 Grenoble, France}
\author{Paulo V. Santos}
\affiliation{Paul-Drude-Institut f{\"u}r Festk{\"o}rperelektronik, 
Leibniz-Institut im Forschungsverbund Berlin e.V., 
Hausvogteiplatz 5-7, 10117 Berlin, Germany}

\title
  {Remotely pumped GHz antibunched emission from single exciton centers in GaAs}

\abbreviations{}
\keywords{impurity centers, excitons, surface acoustic waves, remote pumping, single-photon sources, GHz microwave}

\begin{document}


\begin{abstract}
  Quantum communication networks require on-chip transfer and manipulation of single particles as well as their interconversion to single photons for long-range information exchange. Flying excitons propelled by GHz surface acoustic waves (SAWs) are outstanding messengers to fulfill these requirements. Here, we demonstrate the acoustic manipulation of single exciton centers consisting of individual excitons bound to shallow impurity centers embedded in a semiconductor quantum well. Time-resolved photoluminescence studies show that the emission intensity and energy from these centers oscillate at the SAW frequency of 3.5 GHz. Furthermore, these centers can be remotely pumped  via acoustic transport of flying excitons along a quantum well channel over several microns. Time correlation studies reveal that the centers emit anti-bunched light, thus acting as single-photon sources operating at GHz frequencies. Our results pave the way for the exciton-based on-demand manipulation and on-chip transfer of single excitons at microwave frequencies with a natural photonic interface.
\end{abstract}

\section{Introduction}
Flying photonic qubits are particularly interesting for quantum communication since the photon coherence can be preserved over several kilometers \cite{Marcikic_N421_509_03}. Photons are thus ideal particles for the implementation of quantum functionalities such as entanglement, non-localization and teleportation \cite{Bouwmeester_N390_575_97,Bouwmeester_PL82_1345_99}.  The technical challenges  associated with the manipulation of photonic states are, however, formidable due to the difficulty of bringing two photons into interaction within a short distance. Opto-electronic excitations in the solid state are, in contrast,  much easier to manipulate. Here, \dPVS{the most} promising candidates are flying opto-electronic qubits, which can be used for the exchange of quantum information between remote sites~\cite{Bauerle_RPP81_56503_18}. Recently, flying qubits based on hybrid surface acoustic wave (SAW) structures on semiconductor platforms are attracting increasing attention \cite{PVS326}. One prominent advantage of SAWs is the ability to provide mobile strain and piezoelectric potentials to modulate, confine, and transfer particles between remote on-chip locations. Researchers have coupled SAWs to a variety of systems, including, for example, self-assembled quantum dots \cite{Boedefeld01a,Gell_APL93_081115_08,Metcalfe_PRL105_37401_10}, nanowires \cite{Kinzel_NL11_1512_11,Weiss_NL14_2256_14}, electrons \cite{Hermelin_N477_435_11,McNeil_N477_439_11,PVS321}, superconducting qubits \cite{Gustafsson_S346_207_14,Satzinger_N563_661_18}, diamond NV centers \cite{Golter_PRL116_143602_16,Golter_PRX6_41060_16}, defect centers in SiC \cite{Whiteley_NP15_490_19}, and excitons \cite{PVS177,PVS260,PVS266,PVS279}.  SAWs of $\mu$m-sized wavelengths have also been used to populate two-level systems with charged carriers as well as to  induce the emission of  anti-bunched photons at microwave frequencies \cite{PVS218,PVS246,Hsiao_NC11_917_20}. 
\aPVS{Finally, it has recently been demonstrated that SAWs can effectively manipulate and transport spin-polarized carriers  in semiconductor structures\cite{PVS110,PVS234,PVS240} down to the single particle level,\cite{Bertrand_NN11_672_16,Jadot2021} a further prerequisite for quantum information processing using flying particles.}

\begin{figure*}[!tbhp]
\includegraphics[width=0.85\textwidth]{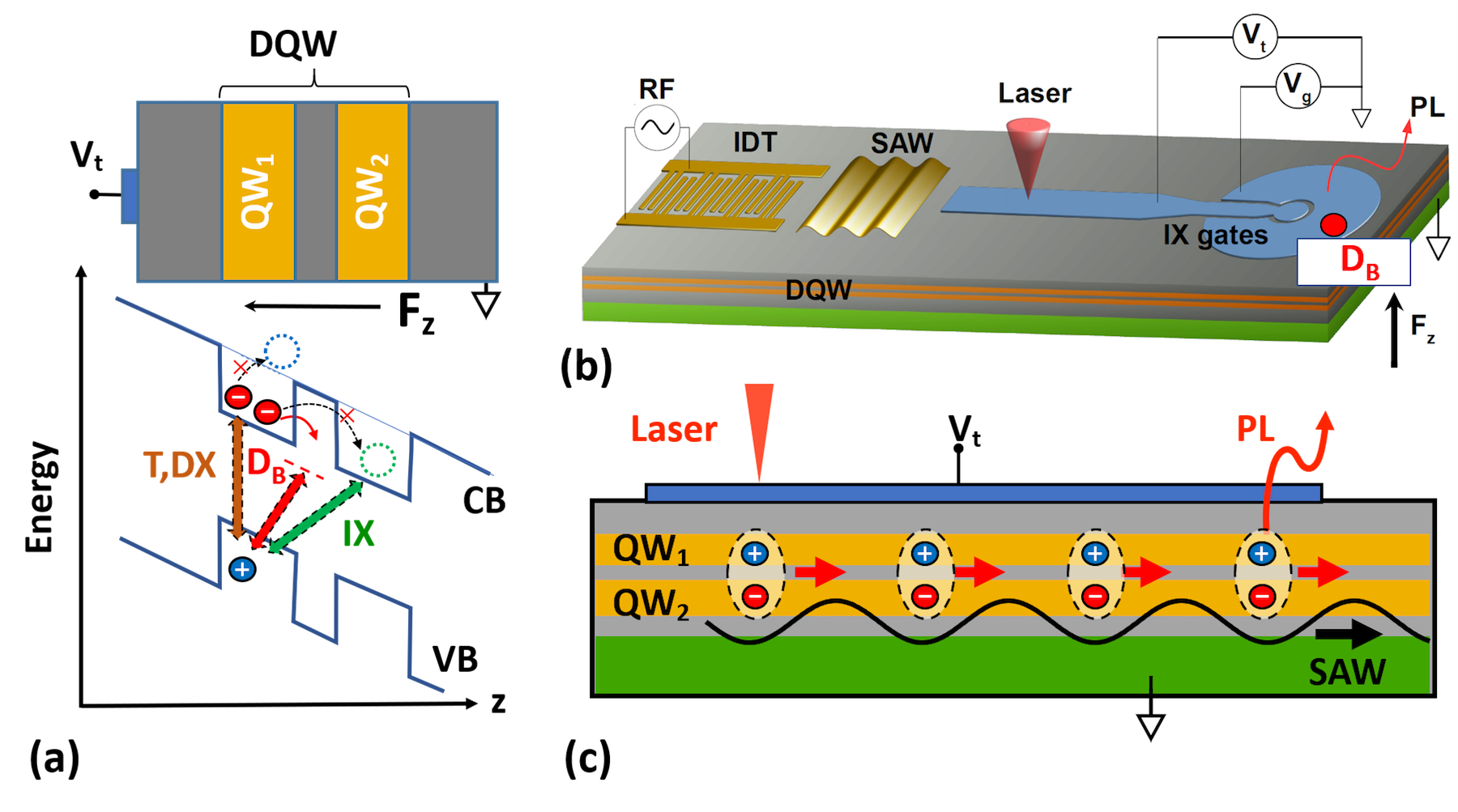}
\caption{{\bf Acoustic manipulations and transport of indirect excitons (IXs)}. 
(a) IX formation via the dissociation of direct excitons (DXs) or trions (Ts) by the electric\aMY{-}field ($F_z$) induced electron tunneling between the quantum wells (QWs) of a double quantum well (DQW). \rPVS{For appropriate $F_z$ values, the T or DX dissociation into IXs (dashed circles) can be blocked, leading to the selective excitation of bound exciton centers ($\DB$).}{Within a narrow field range ($\sim 1$~kV/cm), the trion dissociation into IXs (dashed circles) is blocked leading to the selective excitation of bound exciton states ($\DB$).(Dissociation of a DX is also possible but less favourable energetically.)}
(b)\aMY{ and  (c)} Sketch\aMY{s} of the samples for acoustic IX transport. The IX are formed in the DQW regions underneath a stripe-like semitransparent gate subjected to a bias $V_t$, which enables both electrical control and optical access to the DQW. The stripe\dMY{s} ends in a small circular trap area with a guard gate biased by $V_g$. An interdigit{al} transducer (IDT) launches a SAW, which captures and transports the optically excited IXs along the stripe. The IX distribution is probed by collecting the spatially resolved photoluminescence (PL) along the path.  
}
\label{IXFigSetup}
\end{figure*}

Flying excitons, with their natural inter-conversion to photons, offer several advantages for opto-electronic control as well as for interfacing electronic and photonic excitations. 
Especially suitable for \rMY{the opto-electronic}{ these} applications are the long-living spatially indirect (or dipolar) excitons (IXs) in a double quantum well (DQW) structure subjected to a transverse electric field $F_z$ [cf.~Fig~\ref{IXFigSetup}(a)]. These excitons are formed by the Coulomb binding of electrons and holes driven to different quantum wells (QWs) by the applied field, which controls both the lifetime and the emission energy of the IXs via the quantum confined Stark effect. 
\aPVS{Single IXs can be isolated in small potential traps,\cite{Schinner_PRL110_127403_13} as well as in single defect centers \cite{PVS314} that act as sources of single-photons. }
Analogously to the \rPVS{acoustic transport of charged particles by the SAW piezoelectric field,\cite{PVS156}}{piezoelectric transport of charged particles, \cite{PVS156} } the charge-neutral IXs have a long lifetime and can be confined and transported by the mobile band-gap modulation produced by the SAW strain field,\cite{PVS177} as illustrated in Figs.~\ref{IXFigSetup}(b) and \ref{IXFigSetup}(c). 
%
\aPVS{These interesting properties make flying IXs propelled by SAW fields potential candidates { to complement the functionalities of }  single electrons in nano-electronic circuits requiring an interface to photons.} 
\dPVS{The long-range transport of IXs enabled by their long lifetime has so far only been demonstrated  in wide transport channels using SAWs with wavelengths of a few $\mu$m \cite {PVS177,PVS260,PVS266}. }

A main challenge for the implementation of flying excitonic qubits is the creation of  two-level excitonic states interconnected by a transport channel, which can store single particles and convert them to photons. 
\aPVS{The long-range transport of IXs enabled by their long lifetime has so far only been demonstrated  in wide transport channels using SAWs with wavelengths of a few $\mu$m \cite {PVS177,PVS260,PVS266}. }
In this work, we realize a major step towards this goal by demonstrating the manipulation and remote pumping of single exciton centers by flying IXs propelled by GHz-SAWs in a GaAs-based semiconductor platform. The single centers used here consist of excitons bound  to single shallow impurities (denoted as D$_\mathrm{B}$) in a DQW structure. We have recently reported that these centers can be spectrally isolated and resonantly excited by appropriately biasing the DQW structure~\cite{PVS314}.  Here, we demonstrate the pumping of individual D$_\mathrm{B}$ centers by IXs driven along a narrow transport channel by a SAW. The oscillating SAW strain field modulates the narrow emission lines of the D$_\mathrm{B}$ centers, which can be used as a sensitive probe of the local strain amplitudes. Time-resolved spectroscopic studies show that the recombination lifetime of the   D$_\mathrm{B}$ states is sufficiently short to follow the  3.5~GHz SAW pumping rate. More importantly,  photon correlation investigations reveal that the acoustic pumping of these centers is followed by the emission of anti-bunched photons with a repetition rate corresponding to the SAW frequency, which shows that the center acts as a single photon source operating at very high frequencies.

\dMY{7}
\section{Results and discussion}

\subsection*{Exciton energy modulation by SAWs}

\begin{figure*}[t!bhp]
    \includegraphics[width=0.9\textwidth]{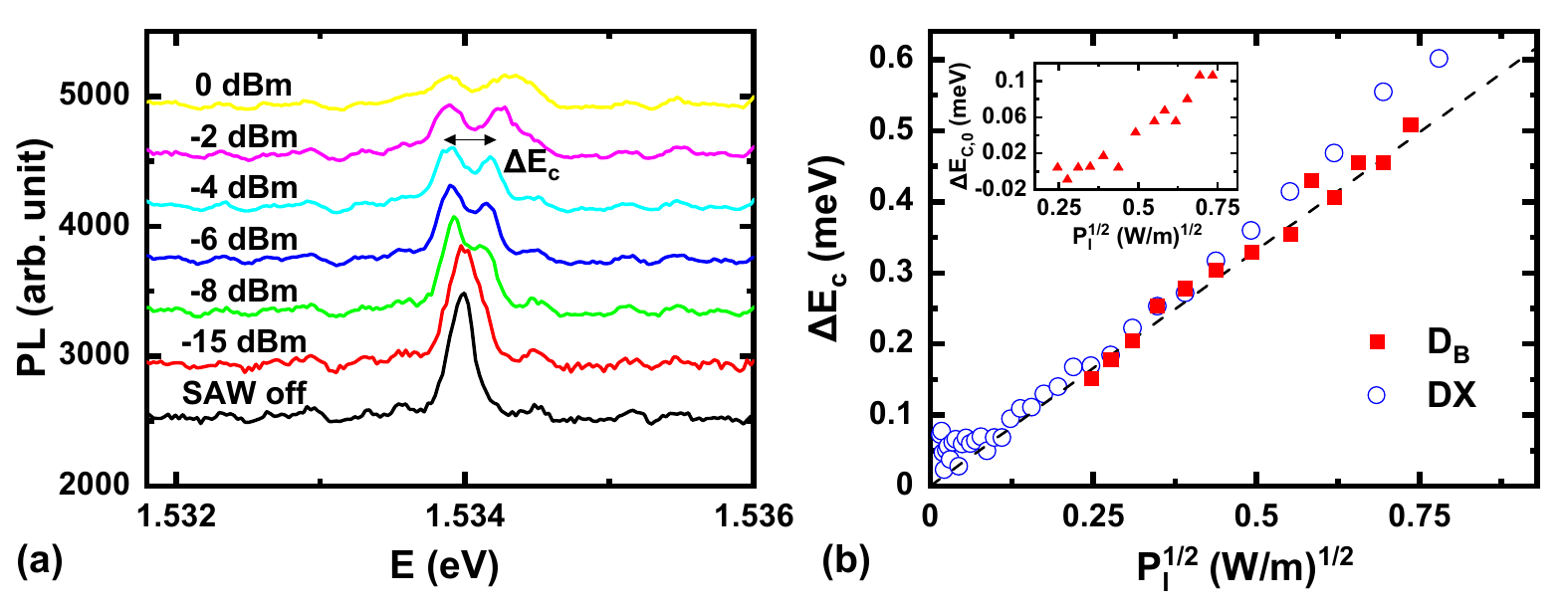}
    \caption{{\bf Acoustic energy modulation of D$_{\rm\bf B}$ \rMY{centers}{ states}} (a) Time-averaged photoluminescence (PL) spectra recorded for different acoustic \rMY{power under local excitation.}{ amplitudes (the dashed lines are guides to the eye). The peak around 1.5355 eV visible at high powers is associated with IX species.} (b) Transition energy modulation amplitudes 
    $\Delta E_\mathrm{B}$ (red squares) and $\Delta E_\mathrm{DX}$ (circles) 
    as a function of the SAW power density. The dashed line yields the corresponding modulation determined from the model in the Supporting Information Sec.~S1. \aMY{Inset: power dependence of the central energy shift, $\Delta E_\mathrm{C,0}$.}
    }
\label{IXFigFWHM}
\end{figure*}

The $\DB$ centers can be resonantly activated under weak optical excitation by biasing  the DQW structure with a voltage $V_t$  close to the onset of IX formation~\cite{PVS314}. Under these conditions, the photoexcited electron-hole pairs bind to free residual carriers to form trions (Ts). The conversion of trions to IXs via electron tunneling through the DQW barrier requires the excitation of a free electron to the band states. As illustrated in Fig.~\ref{IXFigSetup}(a) the tunneling to single D$_{\rm B}$ becomes energetically favorable to the IX formation. The emission from these centers is characterized by a narrow line [with full-width-at-half-maximum (FWHM) of \rMY{0.2}{0.25}~meV, cf. lowest spectrum in Figure~\ref{IXFigFWHM}(a)] spectrally isolated from the IX and direct exciton (DX, excitons whose electron and hole reside in the same QW) transitions.

The spectroscopic studies were carried out \rMY{with}{using} the setup of Figs.~\ref{IXFigSetup}(b) and \ref{IXFigSetup}(c) using SAWs with a wavelength of $\lSAW=800$~nm (corresponding to a frequency of $\fSAW=3.58$~GHz). The photoluminescence (PL) spectra of Fig.~\ref{IXFigFWHM}(a) show that  under an increasing SAW field the $\DB$ line initially broadens and eventually splits into two. \aMY{Here, the laser excitation spot coincided with the location of the $\DB$ center within the spatial resolution (i.e. local excitation).} The SAW strain field periodically modulates the excitonic transition energies $E_\mathrm{C}(t)$ (C = DX, T, $\DB$)  according to:\cite{PVS107}

\begin{equation}
    E_\mathrm{C}(t)=E_\mathrm{C,0}+\frac{\Delta E_\mathrm{C}}{2}\sin\left(\frac{2\pi}{\TSAW}t\right) ,
    \label{EqEc}
\end{equation}

\noindent where   \(\Delta E_\mathrm{C}\) is the peak-to-peak modulation amplitude and \(\TSAW=1/\fSAW \)  the SAW period.

For energy shifts $\Delta E_\mathrm{C}$ smaller than the linewidth, the modulation manifests itself as an apparent broadening of the time-integrated PL lines. For larger modulation amplitudes, the time-averaged PL develops a camel-like shape with peaks at energies $E_\mathrm{C,0}\pm \Delta E_\mathrm{C}/2$ corresponding to the maximum and minimum band-gaps under the SAW field, thus, \rMY{leading to}{ reproducing} the behavior observed in Fig.~\ref{IXFigFWHM}(a). \aMY{At sufficiently high power the piezoelectric field from the SAW quenches the PL, { as} evident in the top curves in Fig.~\ref{IXFigFWHM}(a).} Figure~\ref{IXFigFWHM}(b) displays the dependence of the peak-to-peak modulation amplitude for the $\DB$ ($\Delta E_\mathrm{B}$) and DX ($\Delta E_\mathrm{DX}$) transitions determined from  fits of the measured spectra  to a model for the time-integrated PL line shape under a SAW described in detail in the Supporting Information (SI) Sec.~S2. The dashed line yields the corresponding strain-induced band-gap modulation determined using the GaAs deformation potentials and the  SAW fields in the DQW calculated from the applied rf-power (cf.~Sec.~S1). As expected for a shallow center, the D$_{\rm B}$ energy modulation amplitude follows closely the one for the DX states, which increases proportionally to the SAW amplitude (and, thus, to $\sqrt{\Pl}$, $\Pl$ being the linear power density of the SAW). It is worthwhile to emphasize  that \aMY{the }narrow \aMY{linewidth of} D$_{\rm B}$ \dMY{linewidths of the bound exciton states} enables the quantitative determination of very small strain levels.


\aPVS{The central energy $E_\mathrm{C,0}$ of the SAW-splitted $\DB$ line in Fig.~\ref{IXFigFWHM}(a) also slightly blueshifts with increasing acoustic powers, as illustrated in the inset of Fig.~\ref{IXFigFWHM}(b). The shift $\Delta E_\mathrm{C,0}$ is attributed to changes in the static field $F_z$ caused by charge redistribution within the structure by the SAW piezoelectric field. This effect is nevertheless smaller compared to the splitting $\Delta E_\mathrm{B}$ caused by the SAW strain fields [see Figure~\ref{IXFigFWHM}(b) inset].} 

\subsection*{Long-range \rMY{indirect exciton }{IX} transport}
\label{SI_TRANSPORT}

\begin{figure*}[tbhp]
    \includegraphics[width=0.95\textwidth]{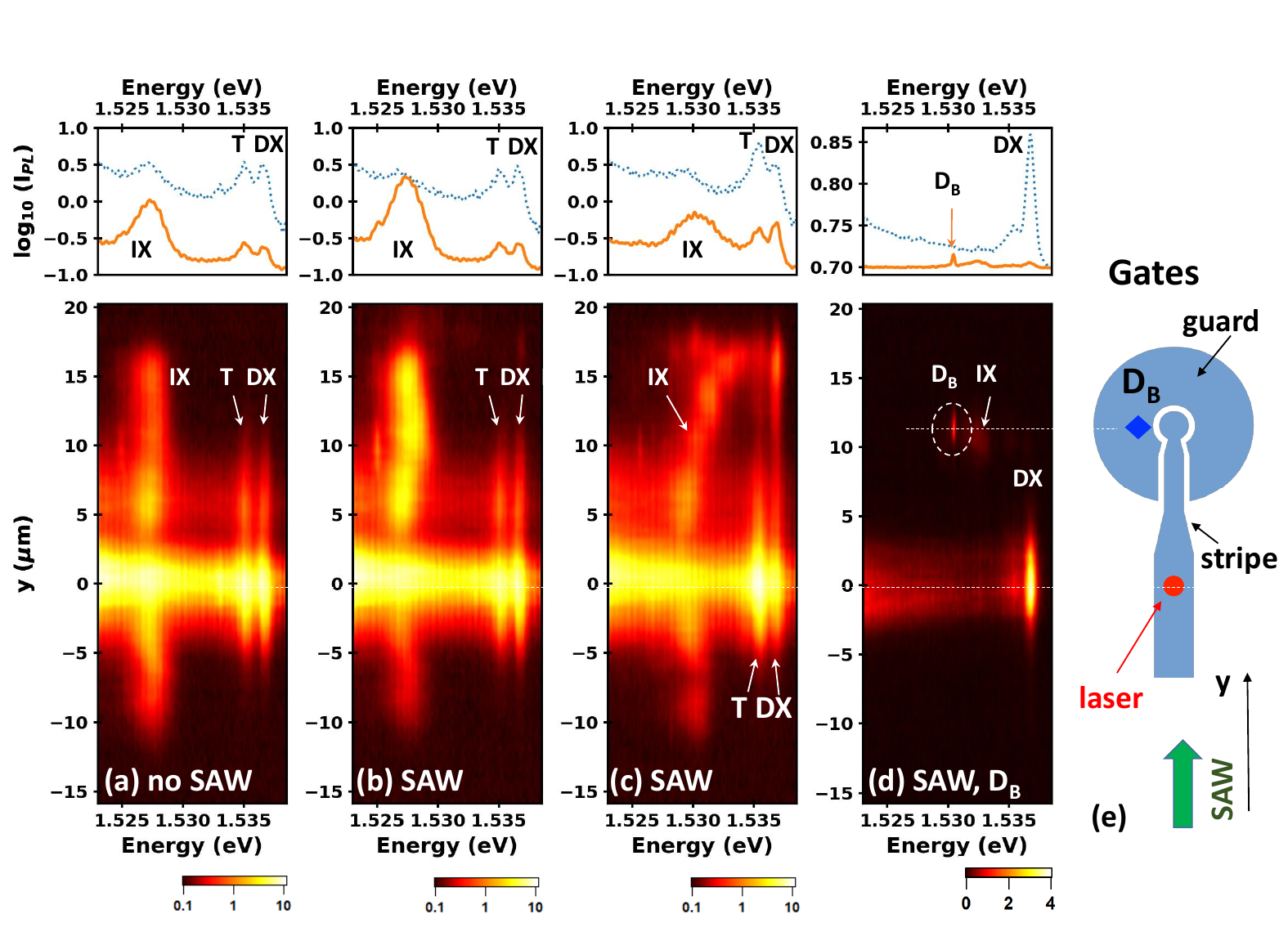}
    \caption{ {\bf Optically detected acoustic transport of IXs.}  (Lower panels) Spatially resolved photoluminescence maps on a log intensity scale in the absence (a) and in the presence of a SAW for (b) $\Prf=-9$~dBm and  $V_t=V_g=0.2$~V and (c) $\Prf=-11$~dBm, for $V_t=0.23$~V, and $V_g=0.31$~V. (d) Corresponding map \aMY{(linear color scale)} with an impurity center $\DB$ on the transport path and biasing conditions to enhance the impurity PL ($\Prf=-14~$dBm and $V_t=V_g=0.4$~V). (\rMY{e}{d}) Sketch of the transport path defined by electrostatic gates. The upper panels in (a)-(d) display profiles of the PL intensity integrated around the excitation region ($|y|<3~\mu$m\aMY{, blue dotted line}) and along the transport path ($y>3~\mu$m\aMY{, orange solid line}) \aMY{on a log scale}.
    }
    \label{IXFigTrans}
    \end{figure*}

The lower panel of Fig.~\ref{IXFigTrans}(a) displays a spectrally resolved PL map of the exciton distribution under the electrostatic stripe gates under optical excitation by a focused laser spot  at $y=0$ [cf. sketch of Fig.~\ref{IXFigTrans}(e)]. This map was recorded in the absence of a SAW under a transverse field $F_z=5$~kV/cm \aPVS{applied} across the DQW. 
The PL around the excitation spot (blue dotted line in the upper panel, integrated for $|y|<3~\mu$m) shows the characteristic emission line\aMY{s} from \aPVS{the} DX\aMY{s}, T\aMY{s}, and IXs   superimposed on a broad PL background from the doped layers and emission centers  in  the substrate (note that the intensity of the  DX and T lines becomes strongly suppressed under  the applied transverse field). Away from the generation area the PL becomes dominated by the emission from IXs, which, due to the long recombination lifetime, can diffuse up to the top region of the guard gate (thick orange line). 
In fact, most of the remote PL from  DXs and Ts arises not from the diffusion  of these species but rather from the conversion of diffusing IXs to DXs or Ts. 
Note also that the diffusing IXs can easily cross the narrow gap between the stripe and guard gate at $y=13~\mu$m.

 Figure~\ref{IXFigTrans}(b) displays a PL map recorded under the same conditions as in Fig.~\ref{IXFigTrans}(a), but now under a  SAW propagating along the $y$ direction. The acoustic field pushes the IXs upwards leading to a strong increase of the IX PL for positive $y$ (cf.~orange line in the upper panel) and a reduction for negative $y$. The recombination energy and location of the transported IXs can be controlled by changing the bias applied to the gates. As an example, Fig.~\ref{IXFigTrans}(c) shows a map recorded by increasing the guard bias by $0.08$~V relative to $V_t$. The IX emission energy blueshifts as IXs enter the guard gate \aMY{due to the reduced $F_z$ underneath the guard gate.} \rMY{Afterwards, they}{as well as  when the particles} are pushed by the SAW beyond  the guard, where they  become converted to DXs or trions. The additional energy \rMY{required to overcome the barrier between DXs (or Ts) and IXs}{for the  blueshift} is provided by the moving SAW field. 

The IX transport  over tens of $\lSAW$ can  remotely activate $\DB$, as shown in Fig.~\ref{IXFigTrans}(d). Here, the SAW amplitude and gate biasing conditions were selected to enhance the emission of a $\DB$  center under the guard gate approx. \rMY{12 }{13}~$\mu$m (corresponding to \rMY{15 }{16}$\lSAW$) away from the excitation spot (dashed circle).

\begin{figure*}[tbhp]
    \includegraphics[width=0.9\textwidth]{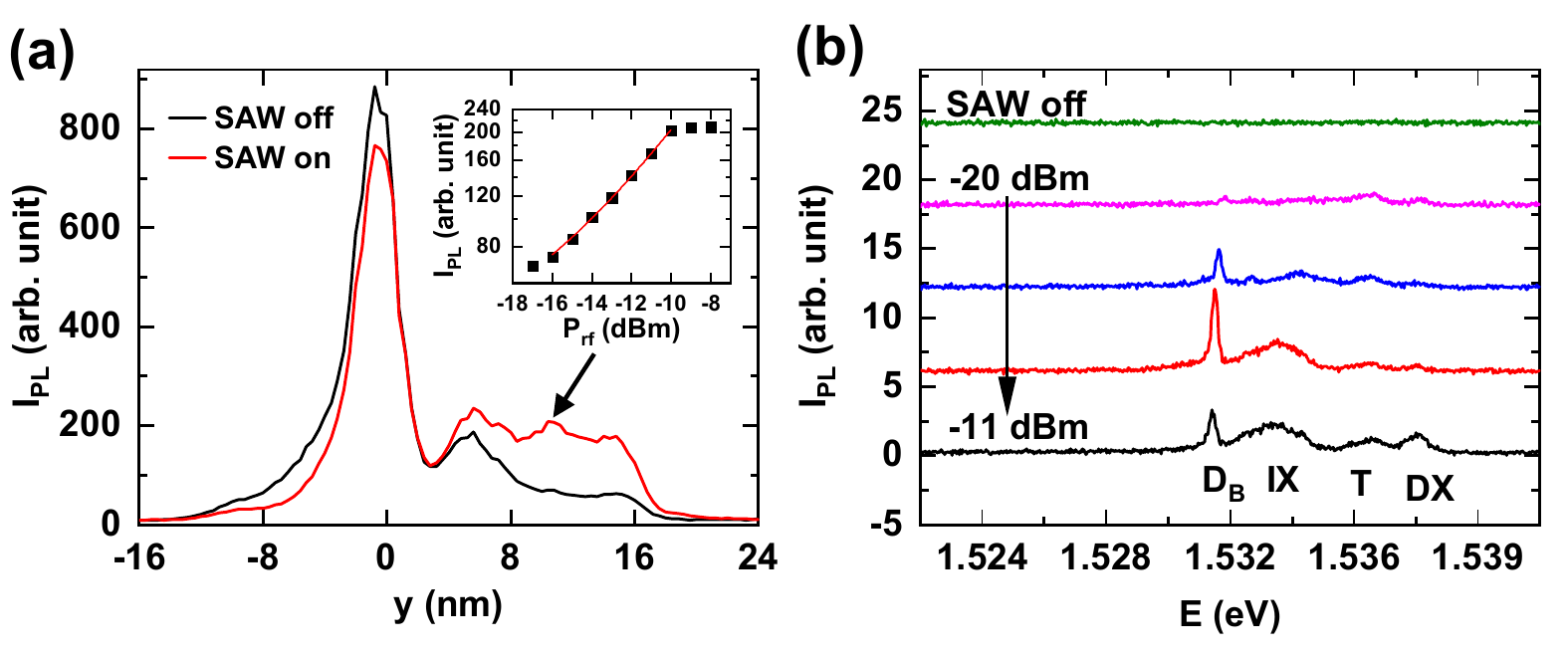}
    \caption{ 
    \aPVS{
      {\bf Transport profiles and their power dependence.} 
      (a) Energy-integrated cross-sections of the IX along the transport ($y$) direction in the absence [black, cf. Fig.~\ref{IXFigTrans}(a)] and under a SAW [red, cf.~Fig.~\ref{IXFigTrans}(b)]. The profiles were obtained by integrating the emission  over the IX energy range (from 1.5265~eV to 1.5308~eV). The acoustic transport enhances the IX PL intensity in the $+y$ direction and reduces it in the opposite direction. Inset: SAW power dependence of the remote IX PL (symbols) together with a linear fit (line). The  PL was recorded at the position $y=10.8$~$\mu$m indicated by the arrow in the main plot. 
      (b) PL spectra recorded for different acoustic powers $\Prf$ 
      around the  remotely activated $\DB$ at $y=12$~$\mu$m in Fig.~\ref{IXFigTrans}(d). The PL intensity of the $\DB$ is first enhanced and then suppressed with increasing SAW power.
      }
    }
    \label{IXFigTrPdep}
    \end{figure*}

\aPVS{A more detailed analysis of the IX SAW transport is presented in Fig.~\ref{IXFigTrPdep}(a). The black and red curves in this figure compare PL profiles along  transport path ($y$) recorded in the absence and presence of a SAW, respectively. In the absence of a SAW, the PL away from the excitation spot  is attributed to the diffusion of the long living IXs. The acoustic transport along $+y$ significantly enhances the IX PL   for $y>8~\mu$m  and reduces it for $y<0$. 
The inset of Fig.~\ref{IXFigTrPdep}(a) displays the  dependence of the remote IX PL  on  SAW power ($\Prf$). Here, the symbols correspond to the PL intensity recorded at the position  indicated by the arrow in the main plot. The IX PL increase due to acoustic transport  only becomes significant for $\Prf$ above $-16$~dBm and saturates above $-10$~dBm. Within this transport window, the PL intensity rises approximately linearly with SAW power, as shown by the line superimposed on the data points. }

\aMY{We now turn our attention to the dependence of the remote $\DB$ emission on SAW power [(cf. Fig.~\ref{IXFigTrPdep}(b)]. Here, the PL spectra were recorded by integrating the emission  over a range $\Delta y=6.4$~$\mu$m  around the $\DB$ at  $y=12$~$\mu$m in Fig.~\ref{IXFigTrans}(d). In addition to the $\DB$ line, the spectra also show emission features associated with IX, T, and the DX species. As in Fig.~\ref{IXFigTrans}(d), most of the trions and DXs contributing to the spectra result not from acoustic transport but rather from the conversion from IXs. 
Interestingly, the remotely excited PL intensity from the $\DB$ center first increases until it reaches a level comparable to the one observed for local excitation, and then becomes suppressed for higher SAW powers. This behavior indicates that SAWs can not only populate the $\DB$ centers, but also depopulate them after a certain power threshold. 
}

\subsection*{Photoluminescence dynamics and  autocorrelation}

\begin{figure}[tbhp]
    \includegraphics[width=.6\columnwidth]{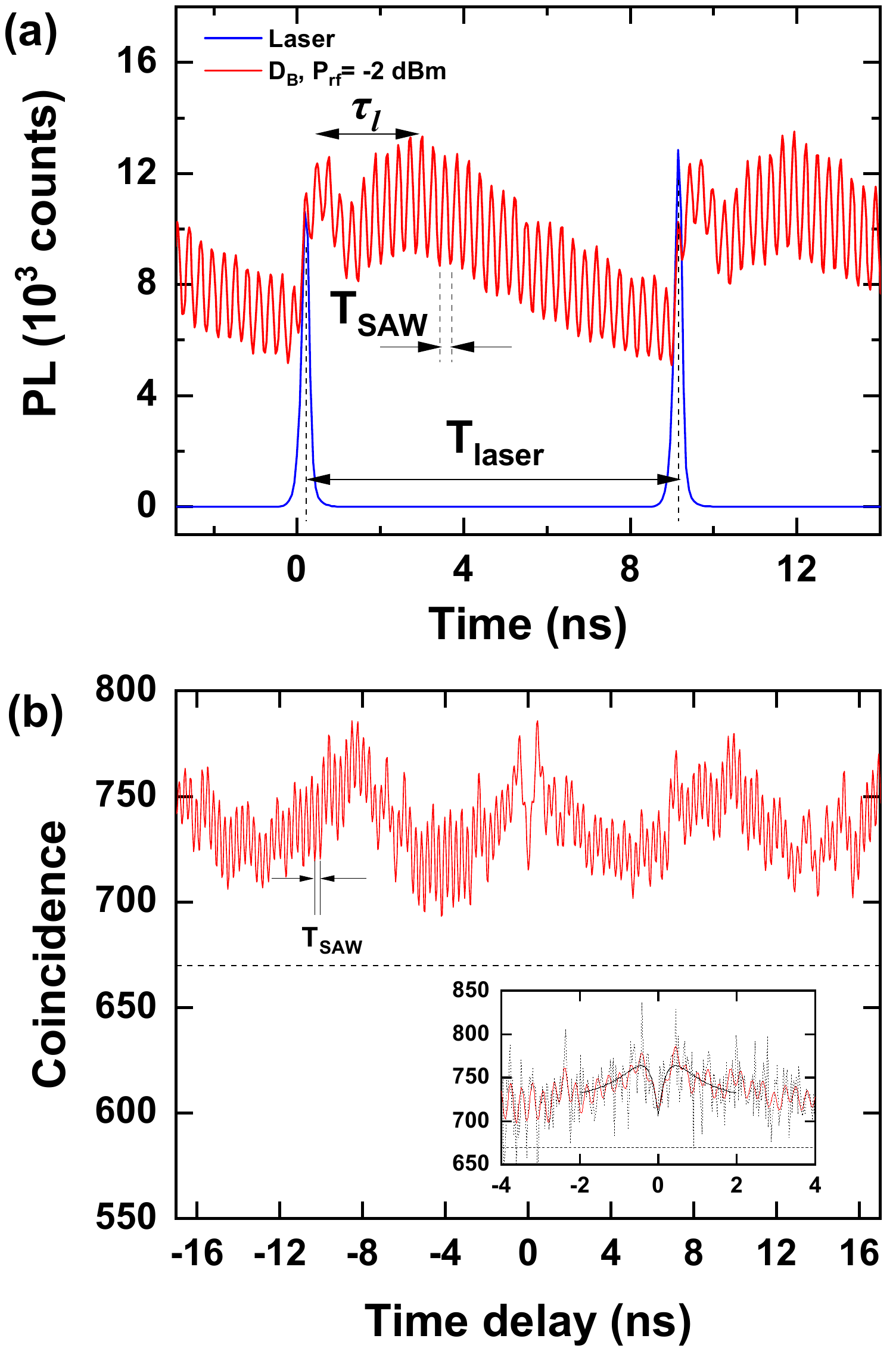}
    \caption{{\bf Emission dynamics and photon autocorrelation from \aMY{remote} single D$_\mathrm{\bf B}$.} 
    \aPVS{(a) Time-resolved  PL  from a $\DB$ center pumped by IX transported by  a $\Prf=-2$~dBm SAW (red). The IXs were photoexcited by a pulsed laser spot  $8~\mu$m away from the $\DB$ center. $\TSAW$ and $T_\mathrm{laser}$ denote the SAW period and laser repetition delay, respectively. $\tau_l$ indicates the SAW propagation time from the excitation spot to the $\DB$. The blue curve displays the corresponding profile for the laser pulses.} 
    \dMY{(a) Time-resolved laser (blue) and PL  emission from a $\DB$ states remotely excited  a $\Prf=-2$~dBm SAW (red). The traces were collected by placing the excitation spot $8~\mu$m away from the $\DB$ center. $\TSAW$ and $T_\mathrm{laser}$ denote the SAW period and laser repetition delay, respectively.}
    \dMY{The inset show a close up of the signal with a time constan t  of 110~ps (red line).}
    (b) Second-order photon autocorrelation $g^{(2)}(\tau_c)$ for the \aMY{same} $\DB$ \rPVS{center}{ state} showing anti-bunching at zero delay \aMY{(FFT filtered)}.  
    \aMY{The horizontal dashed line indicates the recorded background from IX emission close to the $\DB$ center near zero time delay. The inset shows a close-up in the vicinity of zero time delay (dotted line: untreated; red: FFT-filtered) together with a fit to the time-dependence of the anti-bunching feature (black solid line, see text). }
    \dMY{under acoustic excitation ($\Prf=-3$~dBm). The excitation spot was $1~\mu$m away. The dashed envelopes are guides to the eye. The inset shown the averaged ${\bar g}^{(2)}$ around the delays corresponding to multiples of the laser period.}
    }
    \label{IXFigTR}
    \end{figure}

The excitation of the $\DB$ centers by GHz SAW fields induces a strong time modulation of their optical emission.  The red line in Fig.~\ref{IXFigTR}(a) displays the time-resolved PL trace \dMY{of a center} recorded on a $\DB$ center located  about $\ell\sim8$~$\mu$m away from the laser excitation spot. The blue curve reproduces, for comparison, \aMY{the }\dPVS{a} time-resolved profile of the exciting laser spot (not to scale), which  consists of pulses with a FWHM  of about $0.27$~ns and a repetition time of 9~ns. 

The short-period oscillations in the $\DB$ response (red curve) correspond to the SAW period $\TSAW=0.28$~ns \dMY{. A close up of the oscillations (upper inset) reveals that the D$_\mathrm{B}$ emission decays with a time constant  of approx. 110~ps, thus} demonstrating that the PL from these centers can follow the fast varying acoustic field. 
The pulses create a high density cloud of IX \aMY{with long life time}, which partially screen\aMY{s} the \aMY{SAW} modulation potential around the excitation spot. As a consequence, \aMY{within one $\TSAW$, only a fraction of the cloud is captured by the SAW, and} \rMY{the PL}{these} oscillations persist over times longer than the laser repetition period \aMY{of 9~ns, since the IX cloud remains abundant}.
The envelope of the oscillations  features two maxima: the first appears immediately after the laser pulse and the second at a delay of approximately $\tau_{\ell}\sim2.6$~ns (referencing the first maximum).  The first maximum is attributed to the drift of hot excitons propelled by repulsive exciton interaction within the high-density cloud around the excitation spot \cite{PVS279}. 
\dMY{The rising time of about 0.4~ns is attributed to the long lifetime of the IXs.}
 The second maximum is assigned to carriers transported by the SAW, which reach the D$_\mathrm{B}$ after a subsequent delay  $\tau_{\ell}=\ell/v_\mathrm{SAW}$, where  $\vSAW=2960$~m/s is the SAW velocity.

The photon emission statistics of the $\DB$ center\dMY{s} was addressed by recording photon autocorrelation \rMY{$g^{(2)}$ }{($g^2$)} histograms under acoustic excitation using a Hanbury-Brown and Twiss setup. 
\aMY{The experiments were carried out by collecting the PL within a 3 nm spectral band around the $\DB$ emission under a SAW excitation power  $\Prf=-8$~dBm. The excitation laser spot was positioned $\ell\sim8$~$\mu$m away.} 
%
\aPVS{The histogram in the main plot of Fig.~\ref{IXFigTR}(b) displays the autocorrelation data after the application of a fast Fourier transform (FFT) filter {(bandwidths explained in SI Sec.~S3)} to highlight the short modulation period at the SAW period $\TSAW$ (red line): the corresponding untreated (raw) data is shown as a black dotted line in the inset as well as in SI Sec.~S3. As in Fig.~\ref{IXFigTR}(a), the short and long period oscillations in the main plot of Fig.~\ref{IXFigTR}(b) are associated with modulation at the repetition periods $\TSAW$ and $T_\mathrm{laser}$ of the SAW and the laser pulses, respectively. We note that the identification of the short-period SAW oscillations in the raw data (cf. inset) is very challenging since the required accumulation time (or, equivalently, total number of coincidences) increases as $(T_\mathrm{laser}/\TSAW)^2$.  As a consequence, in order to achieve the same level of signal-to-noise ratio as reported in  Ref.~\citenum{PVS218}, where $\TSAW=1.33$ ns and $T_\mathrm{laser}=25$ ns, the accumulation time would need to be 3 times higher for the SAW source investigated  here, where $(T_\mathrm{laser}/\TSAW)^2=32^2$. The SAW induced oscillations become, nevertheless, clearly visible after FFT filtering. We show in  Sec.~S3 that the amplitude of the oscillations at $\TSAW$ is over  20 dB above the noise floor. }


%
%

The histograms of Fig.~\ref{IXFigTR} show a \rMY{distinct }{clear} suppression of the coincidence rate at the time delay $\tau_c=0$ \aMY{, a signature of photon anti-bunching. 
 The black line in the enlarged area presented in the inset  is a fit to the data of an exponential decay function together with a Lorentzian that approximates the   laser  pulse profile (black solid line), from which we extract an autocorrelation recovery time   $\tau_1=0.18\pm0.02$~ns (see details of the fit procedure  in SI Sec.~S3). 
$\tau_1$ is much shorter than the PL decay time of the $\DB$ centers of 0.8~ns  measured in the absence of acoustic excitation (cf. Ref.~\citenum{PVS314}).  The short $\tau_1$ indicates that pumping by high-frequency SAWs significantly reduces the emission time jitter {due to the fast population followed by a fast depopulation of the center by the SAW field, thus reducing the time-interval available for recombination events.}} \aMY{Similarly short response times (between 40 and 300 ps) have also been reported for  GHz single-photon emitters in the form of electrically-driven quantum dots (QDs) \cite{Hargart_APL_102_2013, Schlehahn_APL_108_2016} as well as sources based on SAW-driven electron transport in lateral p-i-n diodes\cite{Hsiao_NC11_917_20}.}

\aMY{The PL collection bandwidth of 3 nm results in a relatively high autocorrelation background caused by IX emission around the $\DB$ center. 
The background level in the coincidence count near $\tau_c=0$ is recorded to be $670\pm40$, as indicated by the horizontal dashed line. 
By taking into account this background correction, we estimate a second-order correlation at zero delay  $g^{2}(0)=0.41\pm 0.21$, i.e., corresponding to an average anti-bunching level below the threshold of $1/2$ for single-photon behavior. This level of anti-bunching is comparable to that reported for 1-GHz SAW-driven electron transport in lateral diodes.\cite{Hsiao_NC11_917_20} The single-photon performance of $\DB$-based sources can be optimized by a better suppression of the spurious IX emission.} 
\dMY{. In order to confirm the selective suppression at $\tau_c=0$, the inset displays the averaged value 
$\bar {g}^{2}(\tau_c)$ for $g^{2}(\tau_c)$ over a time interval of 0.64~ns around time delays multiple   of the laser repetition period. The latter shows that the suppression of coincidences at $\tau_c=0$ is well below the fluctuations, thus  proving the emission of anti-bunched photons. The autocorrelation  $g^{2}(0)=0.75\pm 0.03$, which  corresponds to the simultaneous emission of  4 photons, is probably an upper limit determined by photon collection from the neighboring areas from the center. }


\section*{Discussion}

In conclusion, we have investigated the dynamic modulation and transport of excitons by high-frequency, sub-micron-wavelength SAWs on GaAs DQW structures. In particular, we show that GHz-SAW field can pump single exciton states bound to impurities, which act as two level states emitting anti-bunched photons. 

    The centers can follow the high-frequency acoustic pumping rate leading to the emission of anti-bunched photons synchronized with the SAW phase. The results thus  demonstrate the feasibility of exciton manipulation as well as of exciton-based GHz single-photon sources using acoustic waves\aMY{, the latter being a realization of high repetition rates in single-photon applications}. Finally, multiple single exciton states can be excited by a single SAW beam, thus providing a pathway for scalable arrays of synchronized single-photon emitters. 

\section*{Methods}

\noindent { \bf Sample structure: } The studies were carried out on an (Al,Ga)As DQW  structure consisting of two coupled GaAs QWs grown by molecular beam epitaxy on a n-type doped GaAs(001) substrate [cf.~Fig.~\ref{IXFigSetup}(a)].  The QWs  are  $16$~nm-wide and separated by a 4~nm-thick Al$_{0.33}$Ga$_{0.67}$As barrier.  The electric field $F_z$  induced by the bias $V_{\rm t}$ applied across the structure  drives photoexcited  electrons into  QW$_2$ and holes into QW$_1$, thus increasing the recombination lifetime. Due to the narrow barrier width, the overlap of the electron and hole wavefunctions in the adjacent QWs is still sufficiently strong to maintain the Coulomb correlation required for IX formation.  The DQW structure can thus hold both direct (or intra-QW, DX)  and indirect (inter QW) excitons with transition energies indicated by the brown and green arrows in Fig.~\ref{IXFigSetup}(a), respectively. 

\noindent {\bf Generation of SAWs:}
SAWs with a wavelength of $\lambda_{\rm SAW}=800$~nm (corresponding to a SAW frequency $\fSAW=3.58$~GHz at 4~K) were generated by split-finger aluminum interdigital transducers (IDTs) deposited on the sample surface [cf.~Fig.~\ref{IXFigSetup}(b)]. The depth of the DQW was chosen to yield a type II modulation under the SAW excitation \cite{PVS177}. The IDTs are oriented along a $\langle110\rangle$ surface direction  with a length and width of  150~$\mu$m and 28~$\mu$m, respectively.
The SAW intensity is quantified in terms of either the nominal radio-frequency (rf) power applied to the IDT ($\Prf$) or the SAW linear power density $\Pl$, which is defined as the ratio between the acoustic power and the width of the SAW beam. The latter is obtained by using the measured rf-scattering parameters of the IDTs to determine the fraction of the input rf-power coupled to the acoustic mode. 
 
\noindent {\bf Electrostatic channels for IX transport:}
The IX acoustic transport channel is defined by a semi-transparent Ti stripe placed on the SAW path and biased with a voltage $V_t$ with respect to the doped substrate (cf.~cross-section diagram of Fig.~\ref{IXFigSetup}). The stripe is 2~$\mu$m wide and ends on a small trap (diameter of 0.9~$\mu$m) surrounded by a guard gate with an external diameter of 7.5~$\mu$m. The guard gate, which is biased by a separate voltage $V_g$, reduces lateral stray electric fields in the narrow regions of the stripe, which can dissociate IXs \cite{Schinner_PRB83_165308_11}. As shown in the experiments, IX can easily tunnel over the small separation region (approx.~$0.2~\mu$m) between the stripe and guard gate.

\noindent {\bf Optical spectroscopy:}
Optically detected IX transport experiments were carried out at 4~K with a spatial resolution of approx. $1~\mu$m. The excitons were excited by a spot from a pulsed laser (wavelength of 770~nm, pulse width of 270~ps) focused by a microscope objective on the semitransparent stripe. The photoluminescence (PL) from IXs emitted along the transport path is collected by the same  objective and spectrally analysed by a monochromator \aMY{(double for Fig.~\ref{IXFigFWHM}, single elsewhere.)} with a charge-coupled-device (CCD) detector. Spatially and spectrally resolved PL maps of the IX distribution are obtained by aligning the transport path with the input slit of the spectrometer. The time-resolved PL studies were performed by triggering the laser pulses at a subharmonic ($\fSAW/32$) of the  rf-frequency applied to the IDTs. The PL was in this case spectrally filtered by a \aMY{$810\pm1.5$~nm} band-pass filter \aMY{(Semrock)} and detected  by a superconducting photon detector (Single Quantum) coupled to a time correlator (PicoQuant) with a combined time resolution of 40~ps. \aMY{The data collection time for the autocorrelation measurement was 6 hours.}

\begin{acknowledgement}

The authors thank M. Lopes and A. Hern{\'a}ndez-M{\'i}nguez for helpful discussions and suggestions as well as S. Meister, S. Rauwerdink and A. Tahraoui for the expertise in sample fabrication. S.T. and C.B. acknowledge financial support from the European Union Horizon 2020 research and  innovation program under the Marie Sk\l{}odowska-Curie grant agreement No. 654603 and  JSPS KAKENHI Grant Number {20H02559}.  M.Y., C.B. and P.V.S. acknowledge financial support from the French National Agency (ANR) and Deutsche Forschungsgesellschaft (DFG) in the frame of the the International Project on Collaborative Research SingleEIX Project No. ANR-15-CE24{-0035}/DFG SA-598-12/1.

\end{acknowledgement}




\providecommand{\latin}[1]{#1}
\providecommand*\mcitethebibliography{\thebibliography}
\csname @ifundefined\endcsname{endmcitethebibliography}
  {\let\endmcitethebibliography\endthebibliography}{}

\newpage
\clearpage

\section*{Supporting Information to:}

{\large \bf Remotely Pumped GHz Antibunched Emission from Single Exciton Centers in GaAs}

\vspace{.25cm }

\noindent Mingyun Yuan$^1$, Klaus Biermann$^1$, Shintaro Takada$^2$, Christopher B\"auerle$^3$, and Paulo V. Santos$^1$\\

{\it\noindent $^{1}${Paul-Drude-Institut f{\"u}r Festk{\"o}rperelektronik, 
Leibniz-Institut im Forschungsverbund Berlin e.V., 
Hausvogteiplatz 5-7, 10117 Berlin, Germany}

\noindent $^{2}${National Institute of Advanced Industrial Science and Technology (AIST), National Metrology Institute of Japan (NMIJ), 1-1-1 Umezono, Tsukuba, Ibaraki 305-8563, Japan}

\noindent $^{3}${Univ. Grenoble Alpes, CNRS, Grenoble INP, Institut N{\'e}el, 38000 Grenoble, France}
}\\


\newpage
\beginsupplement

\section{Calculation of the SAW-induced energy modulation}
\label{SI1}

\begin{figure}[tbhp]
\includegraphics[width=0.5\textwidth]{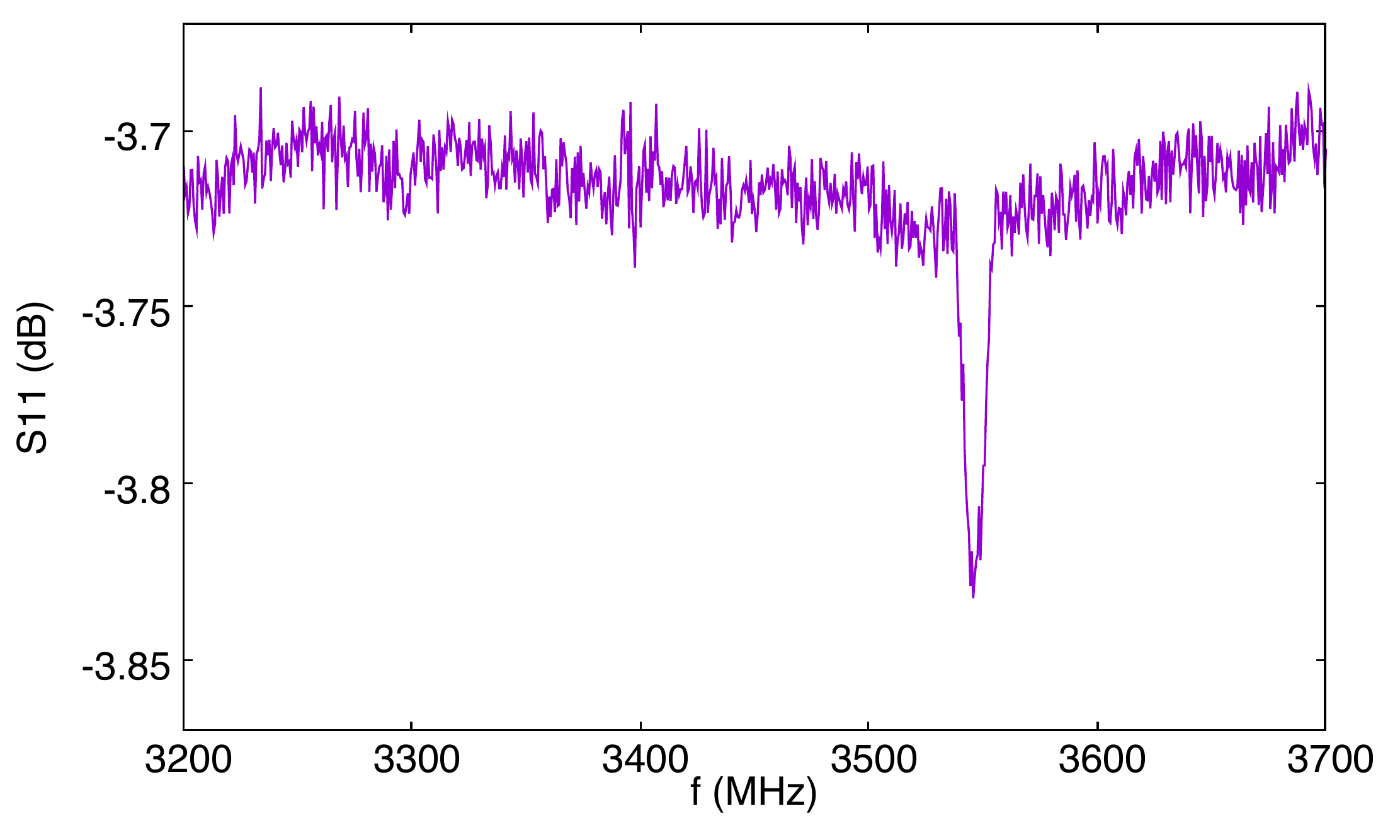}
\caption{Radio-frequency scattering parameter $S_{11}$ (corresponding to the electric reflection coefficient) measured at room temperature for the IDT used to generate the SAWs in the experiments.}
\label{s11}
\end{figure}

\begin{figure}[tbhp]
\includegraphics[width=0.7\textwidth]{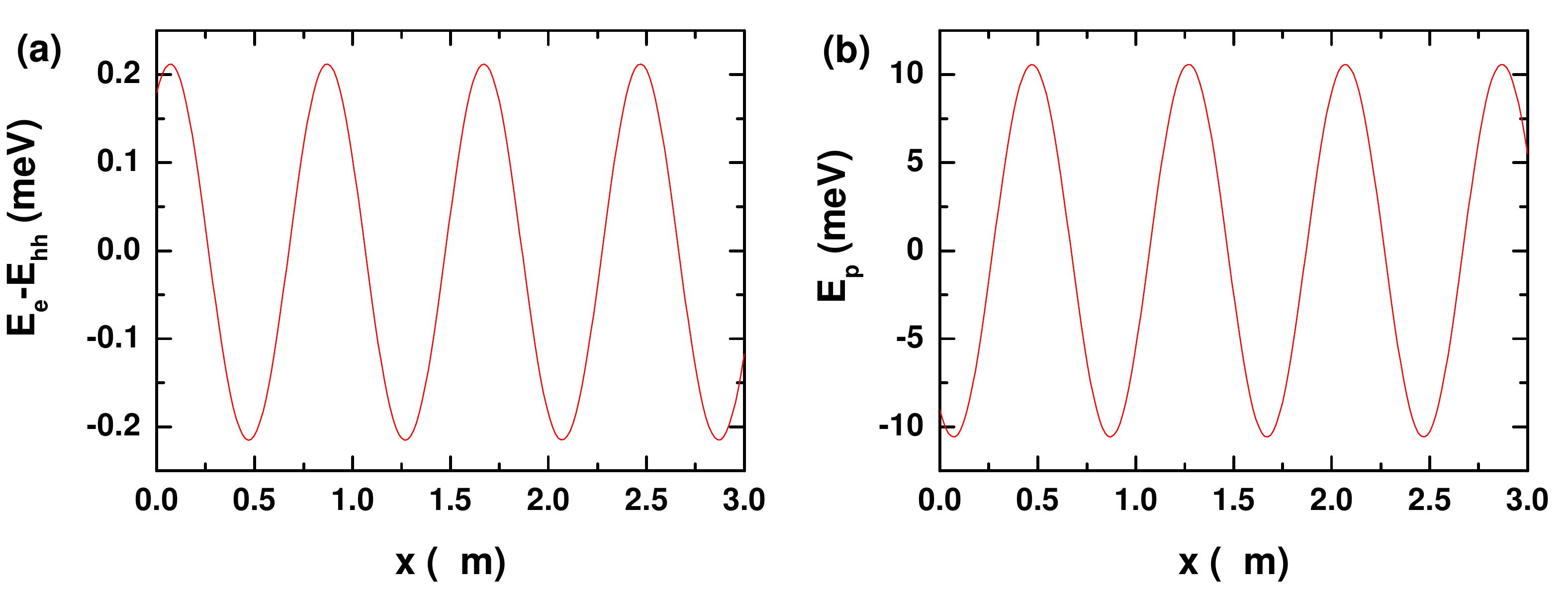}
\caption{(a) Strain-induced bandgap modulation $E_\mathrm{e}-E_\mathrm{hh}$ at the depth of the DQW calculated for $\Pl=0.4$~W/m. (b) The corresponding piezoelectric energy $E_p$.
}
\label{sawmod}
\end{figure}

We calculate in this section the bandgap modulation induced by the SAW. For that purpose, we compare the measured electrical response of the IDTs with numerical calculation of the SAW fields carried out by solving the elasticity equations for the layer structure of the sample~\cite{PVS156}. Fig.~\ref{s11} displays the  rf scattering coefficient $S_{11}$ measured for the IDT, which yields the fraction of the applied rf power reflected by the transducers. From the amplitude of the dip at the resonance frequency $\fSAW=3.54$~GHz, we extract that for a typical applied power of 30~W/m, the transmitted SAW power propagating along one direction of the IDT is $\Pl=0.4$~W/m. The SAW strain modulates the energy of both  the conduction, $E_e$ and the valence bands (heavy hole), $E_{hh}$. The calculated combined bandgap modulation $E_e-E_{hh}$ at the depth of the DQW is shown in Fig.~\ref{sawmod}(a). 
We plot in Fig.~\ref{sawmod}(b) the calculated  piezoelectric energy $E_p$ in the DQW plane underneath the metal gates. For high SAW amplitudes, this field can dissociate excitons, thus inducing a quenching of the  PL. 


\section{Acoustic modulation of the transition energies}
\label{SI_MOD}

\begin{figure*}[!tbhp]
    \includegraphics[width=0.7\textwidth]{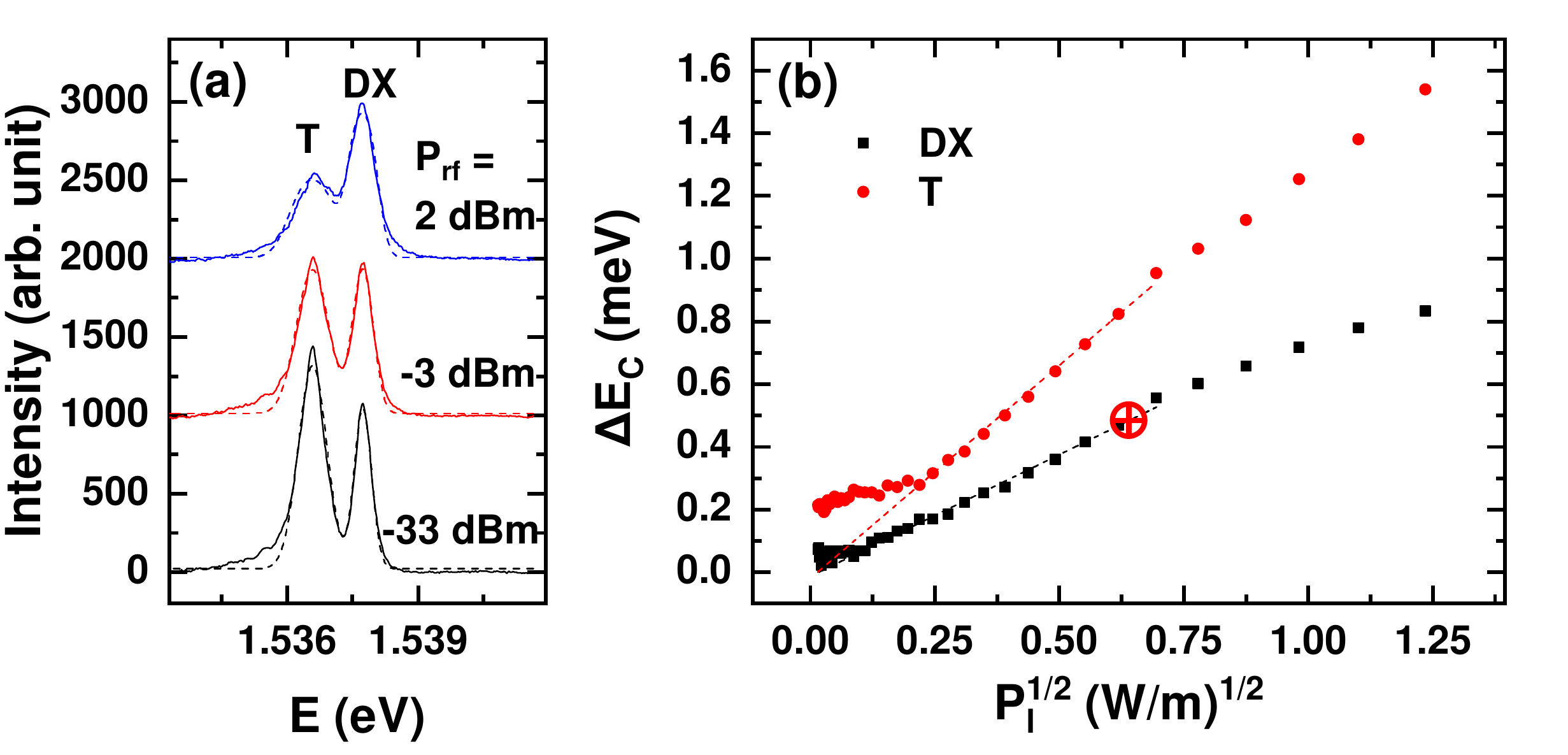}
    \caption{(a) (symbols) Time-averaged photoluminescence (PL) spectra recorded under flat-band conditions for different rf-acoustic powers $\Pl$ applied to the IDT. DX and T denote that direct exciton and trion transitions, respectively. The superimposed curves are fits to Eq.~(\ref{EqIC}). (b) SAW modulation amplitude $\Delta E_C$ (C= DX, T, solid symbols) vs. $\sqrt{\Pl}$. The dashed lines are linear fits for low SAW intensities. 
    }
    \label{free_width_P}
    \end{figure*}
    
In addition to the studies of the dependency of the bound exciton linewidths on acoustic intensity, we also investigated how the acoustic fields impact the DX and trion lines. 
Figure~\ref{free_width_P}(a) displays PL spectra  recorded under  flat-band conditions for increasing SAW intensities 
(quantified by the nominal power $\Prf$ applied to the IDT).  Each spectrum displays two lines associated with the excitation of DXs and trions (T, a DX bound to a free carrier, studied in Ref.~\citenum{PVS314}).  The spectrum for the lowest $\Prf$ essentially corresponds to the PL response in the absence of acoustic excitation. With increasing SAW intensity, both lines slightly broaden and the overall emission intensity decreases.

In order to extract the effects of the acoustic field, we assume that 
PL lines have a Gaussian shape with width $w$ and that their central energy is modulated by the SAW according to Eq.~(1) of the main text. Under these assumptions, the time-integrated PL spectrum can be expressed by the following integral over one SAW period

\begin{equation}
    I_C(E)=
    \frac{1}{\TSAW}\frac{I_{C,0}}{w\sqrt{\pi/2}}\int_0^{\TSAW}\exp\left[-2\left(\frac{E-E_C(t)}{w}\right)^2\right]dt.
    \label{EqIC}
\end{equation}

\noindent For energy shifts $\Delta E_C<<w$, the modulation manifests itself as a broadening of the lines with increasing SAW amplitude. For high modulation amplitudes, the time-averaged PL line splits  into two peaks with energies $E_{C,0}\pm \Delta E_C/2$ corresponding to the maximum and minimum band-gaps under the SAW field.

The lines superimposed on Fig.~\ref{free_width_P}(a) are fits of the measured PL data to Eq.~{\ref{EqIC}}. 
From the fit we extract the SAW modulation amplitudes $\Delta E_C$ for DX and trion,  plotted as symbols in Fig.~\ref{free_width_P}(b) for different SAW amplitudes. The latter is quantified in terms of the SAW linear power density $\Pl$, defined as the acoustic power carried by the SAW mode per unit length perpendicular to the SAW beam. The expected linear dependence on $\sqrt{\Pl}$ is revealed when the SAW-induced apparent broadening exceeds the unperturbed (i.e., in the absence of a SAW) width of the PL line, as shown by the dashed lines. For $\Pl^{1/2}>0.7$ (W/m)$^{1/2}$ one observes a  nonlinear increase of $\Delta E_C$ with SAW amplitude. 

The modulation amplitude determined from the fits for DX (square) correlates well with the band-gap modulation expected from the deformation potential mechanism shown in Fig.~\ref{SI1}.  As an example, the circled red cross in Fig.~\ref{free_width_P}b marks $\Delta E_C$ for $\Pl=0.4$~W/m from the calculation. In contrast, $\Delta E_C$ for trion (circle) deviates significantly from the calculation. It indicates that the simple model of adding the respective modulation of the conduction band and the valance band, used for the calculation here, is not suitable for trion due to the binding to an extra charge.

\section{{Background correction} of the autocorrelation data}
\label{SI3}

\begin{figure}[tbhp]
\includegraphics[width=0.95\textwidth]{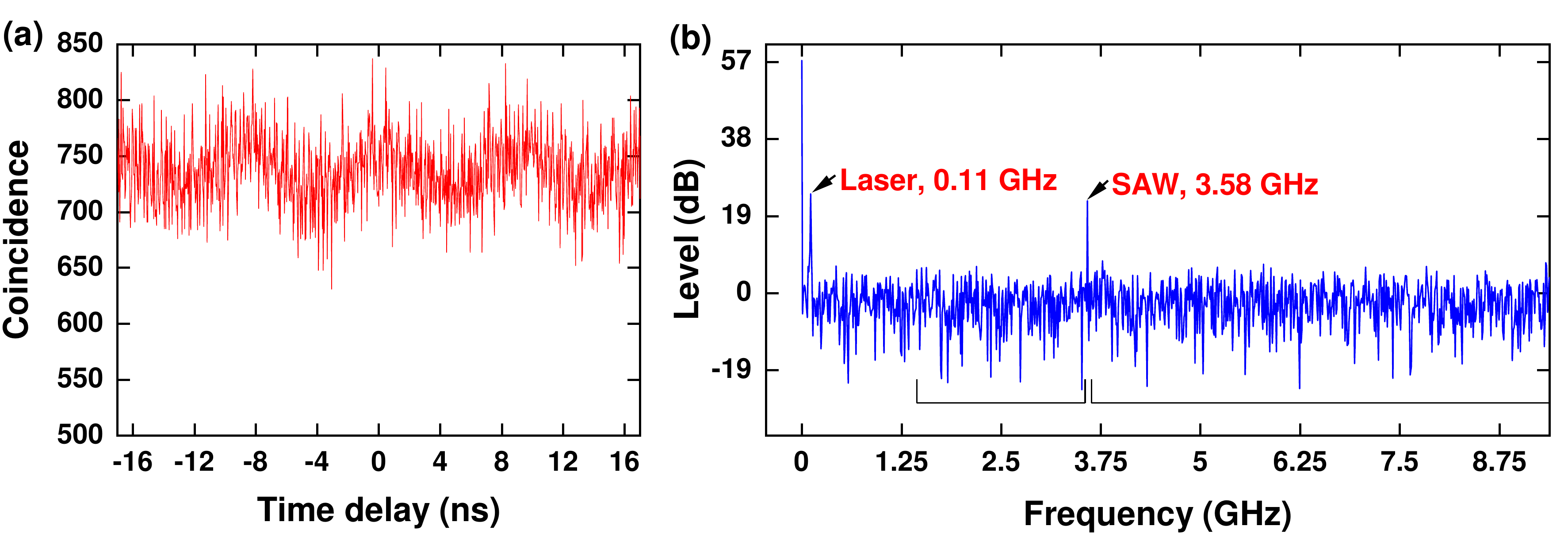}
\caption{{Autocorrelation data processing. (a) Untreated data. (b) FFT spectrum of data presented in (a), showing peaks at the laser frequency and the SAW frequency. The brackets at the bottom indicate the filtered-away bandwidths in Fig. 5(b) of the main text.}}
\label{FFT}
\end{figure}

{In Fig. \ref{FFT}(a) we show the untreated second-order autocorrelation data $g^{(2)}$. Resolving features at frequencies as high as $\fSAW=3.58$ GHz is very challenging for the second-order autocorrelation measurement. To verify the oscillations at $\fSAW$, we perform fast Fourier transform (FFT) to the data and the FFT spectrum is shown in Fig. \ref{FFT}(b). The peak corresponding to $\fSAW$ is very significant, with a signal level more than 20 dB above the noise floor, confirming the SAW modulation. Also at a comparable level is the peak resulting from the laser frequency.}

{Subsequently, we filter away certain bandwidths in the FFT spectrum, indicated by the brackets in Fig. \ref{FFT}(b) to improve the signal-to-noise ratio. \aMY{The time constant $\tau_1$ (see value in the bottom paragraph) of the antibunching feature translates into the frequency of $f_1=1/2\pi\tau_1=0.88$ GHz. We filter out noise above 1.5 times of this frequency with the exception of the narrow bandwidth around the SAW frequency. By choosing this filtering approach we preserve the antibunching feature, resolve the SAW modulation and reject the high-frequency random noise.} The filtered data is presented in Fig. 5(b) of the main text.}

{To estimate the background emission from IXs, we diverted the detection spot away from the $\DB$ center to a position, where the PL from the center disappeared but the IX signal remained similar. A background autocorrelation count $[C_{\rm bgd}]^2$ was recorded at this position. The measured total count $[C_{\rm total}]^2$ can be expressed as $[C_{\rm bgd}+C_{\rm center}]^2$, where $[C_{\rm center}]^2$ is the count from the $\DB$ center that is sought after. From the measurements we extract $[C_{\rm bgd}]^2+2C_{\rm bgd}C_{\rm center}\approx0.85[C_{\rm total}]^2$. The corresponding level for the zero-time delay is indicated as the dashed lines in Fig. 5(b) in the main text.}


{The anti-bunching feature is fitted to a function of the time delay $\tau_c$:} 
\begin{equation}
Coincidence=h_1-b \exp{\left(-\frac{|\tau_c|}{\tau_1}\right)}+h_2\left[\frac{1}{1+(2\tau_c/w)^2}-1\right], 
\end{equation}

{\noindent comprised of an exponential decay with a magnitude of $b$ and a time constant of $\tau_1$, and a Lorentzian with a FWHM of $w$ to approximate the laser-induced pulse profile. The offsets $h_1=780$ and $h_2=60$ are treated as constants. The fitted parameters are $b=74\pm6$, $\tau_1=0.18\pm0.02$ ns and $w=2\pm0.08$ ns.}


\end{document}